\input harvmac.tex
\vskip 2in
\Title{\vbox{\baselineskip12pt
\hbox to \hsize{\hfill}
\hbox to \hsize{\hfill IC/98/224}}}
{\vbox{\centerline{$D=4$ Gauge Theory Correlators From $D=10$ 
NSR $\sigma$-model}
\vskip 0.3in
{\vbox{\centerline{}}}}}
\centerline{Dimitri Polyakov\footnote{$^\dagger$}
{polyakov@ictp.trieste.it}}
\medskip
\centerline{\it The Abdus Salam International Centre for Theoretical Physics}
\centerline{\it Strada Costiera,11 }
\centerline{\it I-34014  Trieste, ITALIA}
\vskip .5in
\centerline {\bf Abstract}
In our previous work (hep-th/9812044)
we have proposed the sigma-model action , conjectured to be 
the NSR analogue of superstring theory on $AdS_5\times{S^5}$.
This sigma-model is the NSR superstring action with
potential term corresponding to the exotic 5-form vertex operator
(branelike state). This 5-form potential plays the role of 
 cosmological term, effectively curving the flat space-time
geometry to that of $AdS_5\times{S^5}$.
In this paper we
study this ansatz in more details and
 provide the derivation of the correlators of
the four-dimensional super Yang-Mills theory  from the above
mentioned sigma-model. In particular,  we show that
the correlation function of two dilaton vertex operators in such a model
reproduces the well-known result of the two-point function in the
$N=4$ four-dimensional super Yang-Mills theory:
$<F^2(x)F^2(y)>\sim{{N^2}\over{|x-y|^8}}$.
\Date{June 99}
\vfill\eject
\lref\myself{D.Polyakov, hep-th/9812044}
\lref\ramf{R.Kallosh, J.Rahmfeld, hep-th/9808038}
\lref\krr{R.Kallosh,J.Rahmfeld, A.Rajaraman,hep-th/9805217}
\lref\mts{R.Metsaev,A.Tseytlin,hep-th/9806095}
\lref\kts{R.Kallosh, A.Tseytlin,{\bf J.High Energy Phys.9810:016,1998}}
\lref\pe{I.Pesando,{\bf JHEP11(1998)002}}
\lref\ampf{S.Gubser,I.Klebanov, A.M.Polyakov, 
{\bf Phys.Lett.B428:105-114}}
\lref\ampt{A.M.Polyakov,hep-th/9809057}
\lref\pol{ J. Polchinski (Santa Barbara, ITP). NST-ITP-99-02, 
 hep-th/9901076 }
\lref\hokerf{E.D'Hoker, D.Freedman, L.Rastrelli,
hep-th/9905049}
\lref\hokers{By E.D'Hoker, D.Freedman, S.Mathur, A.Matusis,
L. Rastelli, MIT-CTP-2843 hep-th/9903196}
\lref\bianchi{M.Bianchi, M.B.Green, S.Kovacs, G.Rossi, JHEP{\bf 08},
013 (1998), hep-th/9807033}
\lref\malda{J.Maldacena, Adv.Theor.Math.Phys.2 (1998)
231-252, hep-th/9711200}
\lref\hokert{D.Freedman, S.Mathur, A.Matusis, L.Rastrelli,
hep-th/980458}
\lref\seiberg{S.Lee, S.Minwalla, M.Rangamani, N.Seiberg,
Adv.Theor.Math.Phys 2, 697 (1998), hep-th/9806074}
\lref\wit{E.Witten{\bf Adv.Theor.Math.Phys.2:253-291,1998}}
\lref\pols{J.Polchinski, L.Susskind, N.Toumbas, hep-th/9903228}
\centerline{\bf Introduction}
One of the most profound questions with regard to critical
superstring theory in ten dimensions is its relation 
to our four-dimensional world.
Superstring theory on $AdS_5\times{S^5}$,
whose low energy limit is the anti de Sitter supergravity,
 is presumed to
be related to $D=4$ $N=4$ super Yang-Mills theory
due to the holography principle and to contain 
crucial information about the large $N$ gauge dynamics 
in four dimensions ~\refs{\malda, \ampf, \ampt, \wit}.
Construction of the action for this theory in the GS formalism
has been considered in ~\refs{\mts, \kts, \pe, \krr, \ramf}.
The GS superstring action in the $AdS_5\times{S^5}$ metric
is essentially non-linear and its 
 quantization in a straightforward way and computation of string scattering
amplitudes seem to be problematic.
In the previous work ~\refs{\myself} we attempted to  formulate the
NSR version of superstring theory on $AdS$.
Namely, we have proposed the $\sigma$-model  action, which we claimede
to be the NSR analogue of the superstring action on 
$AdS_5\times{S^5}$ in the GS formalism.
Up to picture-changing transformation, the action for this sigma-model
 is given by:
\eqn\grav{\eqalign{S_{{NSR}}=\int{d^2}z
\lbrack{1\over2}\partial{X^m}\bar\partial{X_m}
+{1\over2}(
\psi^{m}\bar\partial\psi_{m}+\bar\psi^m\partial\bar{\psi_m})\cr+
\lambda(X)
\epsilon^{p_1p_2p_3p_4}\lbrack{e^\phi}\psi_{p_1}\psi_{p_2}
\psi_{p_3}\psi_{p_4}\psi_t\bar\partial{y^t}+\partial(e^\phi\psi_{p_1}
\psi_{p_2}\psi_{p_3})\bar\partial{{\tilde{x}}_{p_4}}\rbrack+ghosts
\rbrack}}
where $X^m\equiv{(x^p,y^t)}$ are ten-dimensional 
space-time coordinates, split in the $''4+6''$ way,
$p=0,...,3;t=4,...,9$; $\psi$ are worldsheet NSR fermions,
$\phi$ is  bosonized superconformal ghost field.
In other words, the action (1) consists of a kinetic part which is just 
the standard NSR free superstring action and the sigma-model type
potential, determined by the massless 5-form vertex operator,
whose zero-momentum part is given by
\eqn\lowen{V_5=\lambda\epsilon^{p_1...p_4}\psi_{p_1}...\psi_{p_4}\psi_t
\bar\partial{X^t}+ ghosts}
where $\lambda$ is  constant. We will see, from analyzing the BRST
invariance condition that $\lambda$ must infact be constant in the space-time,
i.e. the potential term in the sigma-model is cosmological-like.
The BRST-invariant 5-form massless vertex operator (2) does not
seem to correspond to any known physical particle in perturbative string
spectrum. On the contrary, it is related to the brane dynamics
and space-time geometry
in some very intriguing way. In this paper we will discuss it 
in much more details.  
 The related potential part in the action (1)
is presumed to play the role of  cosmological term,
effectively curving the flat ten-dimensional space-time
to that of $AdS_5\times{S^5}$; the orientation of the 
four-dimensional $AdS_5$ boundary in ten-dimensional
Minkowski space-time is thus related to the polarization of the 5-form
vertex operator (2).
 This note is organized as follows.
We begin with reviewing the arguments that relate the
string theory on $AdS_5\times{S^5}$ to the sigma-model with 
the 5-form state (2).
Then we analyze  the BRST invariance of the five-form vertex operator.
Firstly, we find that the BRST invariance condition for the vertex operator (2)
restricts the dynamics of the scalar field to four dimensions,
longitudinal with respect to the underlying three-brane;
then the condition of the worldsheet conformal invariance further restricts
$\lambda$, requiring it to be constant (which is, in fact, related to $N$).
Then we compute dilatonic correlation functions
in the sigma-model (1) (up to the order of $\lambda^2$) (which, according
to the AdS/CFT correspondents should be related with the correlators
of gauge-invariant $F^2$  operators in the $D=4$ $N=4$ SYM).
   We find, in particular, that the two-point dilaton correlation
function reproduces the $<F^2F^2>$ correlator in the four-dimensional
 Yang-Mills, thus producing the evidence of the connection between
the brane-like sigma-model (1) and the large N limit of gauge theory.

\centerline{\bf NSR Strings on AdS and the brane-like sigma-model}

 Let us briefly recall the arguments of the previous paper that led us 
to relate the sigma-model (1) to the AdS superstrings. 
We start with the $SO(1,3)\times{SO(6)}$-invariant 
$AdS_5\times{S^5}$ action in the GS formalism with fixed kappa-symmetry.
The action is given by ~\refs{\kts}:
\eqn\grav{\eqalign{S={-{1\over2}}\int{d^2}z\lbrace
{\sqrt{-g}}g^{ij}(y^2(\partial_i{x^p}-2i{\bar\theta}{\Gamma^p}\partial_i
{\theta})(\partial_j{x^p}-2i\bar\theta\Gamma^p\partial_j\theta
 )+{1\over{y^2}}{\partial_i}y^t\partial_j{y^t})\cr
+4\epsilon^{ij} \partial_i{y^t}\bar\theta{\Gamma^t}
\partial_j\theta\rbrace }}
On the other hand,
 as the $AdS_5\times{S^5}$ space is 
the near-horizon limit of the D3-brane solution of the type IIB supergravity 
in ten dimensions, one may alternatively view the string theory
on $AdS_5\times{S^5}$ as a theory of closed strings propagating in the $flat$
space-time
in the presence of N parallel D3-branes or, equivalently, 
open Dirichlet strings.
 The stress-energy tensor of a GS superstring in the flat background
is given by:

\eqn\grav{\eqalign{T_{ij}=\Pi^m_i\Pi_{mj}\cr
\Pi^m_i=\partial_i{X^m}+i\bar\theta^\alpha\Gamma^m_{\alpha\beta}
\partial_i\theta^\beta}}

Because of the D3-brane presence the holomorphic and anti-holomorphic spinors
$\theta$ and $\bar\theta$ are no longer independent;
they are related as
\eqn\lowen{{\bar\theta_\alpha}=\lambda(N)(\Gamma^0...\Gamma^3)_
{\alpha\beta}\theta_\beta}
where $\lambda(N)$ is some constant depending on the number N of parallel
D3-branes.
Now in order to find the NSR analogue of the  superstring action in the
presence of $N$ parallel D3-branes, one has to use  the standard relation
between GS space-time spinor $\theta$ and the NSR matter $+$
ghost spin operator of zero conformal dimension:
\eqn\lowen{\theta_\alpha(z)=e^{\phi\over2}\Sigma_\alpha(z)+ghosts}
with the additional constraint (5).
In the absence of the constraint (5) substituting (6) into
the GS expression (4) for the stress-energy tensor would have been
given , up to picture-changing, simply by the stress-energy tensor of  a
free NSR superstring.The constraint (5), however, modifies the 
result of the internal normal reordering that one has to perform
in the terms of the form $\bar\theta\partial\theta\bar\partial{X}$
in the GS stress-energy tensor (4).
Namely, using the relations (5), (6) and the O.P.E. formula between
spin operators: 
$:\Sigma_\alpha(z)\Sigma_\beta(w):\sim{{\epsilon_{\alpha\beta}}\over
{(z-w)^{5\over4}}}+{\sum_p}{{\Gamma^{m_1...m_p}_{\alpha\beta}\psi_{m_1}...
\psi_{m_5}(w)}\over
{(z-w)^{{5\over4}-p}}}$
we find that
\eqn\lowen{:\bar\theta\Gamma^m\partial\theta\partial{X_m}:=
\epsilon^{p_1...p_4}\partial{e^\phi}\psi_{p_1}...\psi_{p_3}\partial{X_{p_4}}
+\epsilon^{p_1...p_4}e^\phi\psi_{p_1}...\psi_{p_4}\psi_t\partial{X^t}+
ghosts}
Other terms in the stress-energy tensor (4) produce, up to picture changing,
the standard matter$+$ghost stress-energy tensor of free NSR theory.
The action corresponding to the 
stress tensor (7) is then the one given by the sigma-model (1).
The above arguments led us to suggest 
that the $AdS_5\times{S^5}$ structure of the the space-time may
in fact be ``depicted'' by the potential term in the sigma-model (1)
  corresponding
to the exotic 5-form branelike vertex (2).
In this paper we will test this conjecture
by computing the correlators in the sigma-model (1) and
showing them to correspond to the four-dimensional correlators in the
large N gauge theory.
First of all we would like to point out at some important properties of the 
vertex operator $V_5$. At zero momentum its form is given by (2).
At a given momentum $k$ the straightforward generalization would be
$V_5(k)=\lambda(k)e^\phi\psi_{p_1}...psi_{p_4}\psi_t
(\partial{X^t}+i(k\psi)\psi^t)e^{ikX}$;
however the BRST invariance imposes constraints on possible values
of $k$. In fact, it restricts $k$ to the  longitudinal directions,
effectively reducing  the dynamics of
 the field $\lambda(k)$ to four dimensions !
The easiest way to see it is to consider the internal singularities
inside the vertex $V_5(k)$.
Recall that if one takes, for instance, the vertex operator of 
a vector boson of the form $e_m(k)(\partial{X^m}+...)e^{ikX}$,
there is the internal singularity due to the coupling
between $\partial{X}$ and  the exponent $e^{ikX}$:
$\partial{X^m}(z)e^{ikX}(w)\sim{{k^m}\over{z-w}}e^{ikX}$.
To remove this internal singularity or, equivalently,
to insure the BRST invariance of the vector vertex operator,
one has to require the transversality of the polarization vector
at non-zero values of the momentum:
$k_me^m(k)=0$.
In case of the vertex operator $V_5$ the 5-form state is BRST invariant
and the internal singularities are absent if the transverse
field $\bar\partial{X^t}$  does not couple to the exponent at all,
i.e. the exponent depends on the longitudinal components of the momentum
only since the O.P.E. $\bar\partial{X^t}(z)e^{ik^{||}X}(w)$ is non-singular
($k^{||}X\equiv{k_p}X^p; p=0,1,2,3$).The same condition can be 
derived in a straightforward way, by computing the commmutator of $V_5(k)$
with the BRST charge.
Therefore the BRST-invariant 5-form vertex operator is purely
longitudinal and the scalar field $\lambda$ is confined to four dimensions:
\eqn\lowen{V_5(k^{||})=\lambda(k^{||})\epsilon^{p_1...p_4}e^{\phi}\psi_{p_1}...
\psi_{p_4}\psi_t(\bar\partial{X^t}+i(k^{||}\bar\psi)\bar\psi^t)
e^{ik^{||}X}}
However, this is not yet the end of the story.
Apart from the BRST invariance of the $V_5$ vertex, we also 
need to insure that the 5-form term in the sigma-model action
(1) does not violate the worldsheet conformal invariance.
Consider the stress tensor:

\eqn\lowen{T(z)={1\over2}(\partial{X^m}\partial{X_m}
+\psi^m\partial\psi_m)+\lambda(x^{||})
\epsilon^{p_1...p_4}e^{\phi}\psi_{p_1}...\psi_{p_4}
\psi_t\partial{X^t}(z)}
The two-point correlation function is given by:
\eqn\lowen{<T(z)T(w)>={1\over{2|z-w|^4}}(15+2<\lambda(x^{||}(z,\bar{z}))
\lambda(x^{||}(w,\bar{w}))>)}
(note that $\partial{X^t}$ doesn't interact with $\lambda$.
It follows that, in order to preserve the conformally invariant
form of the  O.P.E. for
two stress-energy tensors in the two-dimensional CFT
(that is, there are no extra singularities coming from 
the O.P.E. $\lambda(x^{||}(z))\lambda(x^{||}(w))$)
we have to require that $\lambda$  is $x^{||}$-independent,
i.e. it is constant in the four-dimensional longitudinal
subspace.That means that in the momentum space $\lambda(k)$ must
be inversely proportional to the fourth power of $|k^{||}|$:
\eqn\lowen{\lambda(k)\sim{{\lambda_0^{(+1)}}\over{{k^{||}}^4}}}
where $\lambda_0^{(+1)}$ is $k^{||}$-independent.
The label $(+1)$ refers to the ghost number.

\centerline{\bf Correlation functions in the brane-like sigma-model}

In our calculation of the two-point dilaton correlation in the 
sigma-model with the 5-form state we would like to use 
the slightly modified version of (1).Namely, in order to maintain
the correct ghost number balance in correlation functions on the sphere
it is convenient to choose the linear combination of $V_5$
in the $+1$-picture with its picture $-3$
version 
\eqn\lowen{V^{(-3)}={{\lambda_0^{(-3)}}\over{{|k^{||}}^4}}
\epsilon^{p_1...p_4}
e^{-3\phi}\psi_{p_1}...\psi_{p_4}\psi_t
(\bar\partial{X^t}+i(k^{||}\bar\psi)\bar\psi^t)e^{ik^{||}x^{||}}}
where $\lambda^{(-1)}$ and $\lambda^{(-3)}$ are some
$k$-independent constants to be specified later(in fact
both are related to the parameter $N$ of the
gauge theory)
The generating functional for the dilaton amplitude is then given by:
\eqn\grav{\eqalign{Z(\varphi,\lambda)=
\int{D}\lbrack{X}\rbrack{D}\lbrack\Psi\rbrack
\lbrack{ghosts}f(\Gamma)\rbrack{exp}{\lbrace}\int{d^2{z}}
{1\over2}(\partial{X_m}\bar\partial{X^m}+\psi_m\bar\partial\psi^m)
\cr+\int{{{d^4}k^{||}}\over{{k^{||}}^4}}\epsilon^{p_1...p_4}
({\lambda_0^{(-3)}}e^{-3\phi}+{\lambda_0^{(+1)}}e^{\phi})
\psi_{p_1}...\psi_{p_4}\psi_t
(\bar\partial{X^t}+i(k^{||}\psi)\psi^t)e^{ik^{||}x^{}}+cc
\cr+\int{d^{10}k}V_\varphi(k)\varphi(k)\rbrace
}}
where $V(k)$ is the dilaton vertex operator at momentum $k$ and
 $f(\Gamma)$ is certain function of picture-changing operator $\Gamma$,
necessary to insure the correct ghost number balance in correlation 
functions.The function $f(\Gamma)$ will be specified later.
Of course, it is possible to rewrite the functional (13) in an equivalent
form so that it would only contain the picture $+1$ five-form in the
potential, but with the different measure deformation
$f(\Gamma)$.
 We have omitted the term with the full derivative of the three-form,
which does not contribute to correlators in our calculation;
$V_\varphi(k)=\partial{X^m}\bar\partial{X^n}
(\eta_{mn}-k_m{\bar{k}_n}-k_n\bar{k}_m)e^{ikX}$ is the dilaton
vertex operator; in this paper we will ignore the longitudinal
$(k,\bar{k})$ part of the dilaton vertex since it is not important for
the correlation functions.

The two-point dilaton correlation function 
is given by:
\eqn\grav{\eqalign{{<V_\varphi(p_1)V_\varphi(p_2)>}=
{{\delta{Z(\varphi,\lambda)}}\over{\delta\varphi(p_1)\delta\varphi(p_2)}}
|_{\varphi=0}}}
The first non-trivial (of order $\lambda^2$) contribution to this
correlation function
is given by
\eqn\grav{\eqalign{A=<V_\varphi(p_1)V_\varphi(p_2)>\sim
\lambda^{(+1)}\lambda^{(-3)}\int{{d^4{k_1^{||}}}\over{{k_1^{||}}^4}}
\int{{d^4{k_2^{||}}}\over{{k_2^{||}}^4}}
<V_\varphi(p_1)V_\varphi(p_2)V_5(k_1^{||})V_5(k_2^{||})>}}
As we shall observe, the term of order $\lambda^2$ in the expansion
corresponds to the contribution of the $s$-wave of the dilaton
in the AdS picture. As for higher order terms in the expansion
in $\lambda$, we shall argue that they correspond to higher partial waves
of the dilaton field on the $AdS_5$.
Due to the momentum conservation, we have $p_2=-p_1-k_1^{||}-k_2^{||}$,
therefore the $\lambda^2$ part of dilaton amplitude 
is given by:
\eqn\grav{\eqalign{A(p_1)=
{\lambda^{({+1})}}\lambda^{(-3)}\int{{d^4{k_1^{||}}}\over{{k_1^{||}}^4}}
\int{{d^4{k_2^{||}}}\over{{k_2^{||}}^4}}<V_\varphi(p_1)
V_\varphi(-p_1-k_1^{||}-k_2^{||})>=\cr
\int{{d^4{k_1^{||}}}\over{{k_1^{||}}^4}}
\int{{d^4{k_2^{||}}}\over{{k_2^{||}}^4}}\int{d^2}z_1\int{d^2}z_2
\int{d^2}w_1\int{d^2}w_2\lbrace\cr
<e^{-3\phi}:\psi^5_{t_1}:\bar\partial
{X^{t_1}}e^{ik_1^{||}X}(z_1,\bar{z_1})e^{\phi}:\psi^5_{t_2}:\bar\partial
{X^{t_2}}e^{ik_2^{||}X}(z_2,\bar{z_2})
\cr\times{e^{-\bar\phi}}
(\partial{X_{m_1}}+i(p_1\psi)\psi_{m_1})\bar\psi^{m_1}e^{ip_1{X}}
(w_1,{\bar{w}_1})
\cr{\times}e^{-\bar\phi}
(\partial{X_{m_1}}+i(p_2\psi)\psi_{m_1})\bar\psi^{m_1}e^{ip_2{X}}
(w_2,{\bar{w}_2})>\rbrace}}
where we have introduced the notation
$\psi^5_{t}=\epsilon^{p_1p_2p_3p_4}:\psi_{p_1}\psi_{p_2}
\psi_{p_3}\psi_{p_4}\psi_t:$
and the correlators are now to be evaluated in the free theory.
Dividing by the $SL(2,C)$ volume one has:
\eqn\grav{\eqalign{A(p_1)={\lambda^{({+1})}}\lambda^{(-3)}
\int{{d^4{k_1^{||}}}\over{{k_1^{||}}^4}}
\int{{d^4{k_2^{||}}}\over{{k_2^{||}}^4}}
\int{d^2}{w_1}|z_1-w_2|^2|z_2-w_2|^2|z_1-z_2|^2\cr\times
<e^{-3\phi}(z_1)e^{\phi}(z_2)><e^{-\bar\phi}\bar\psi^{m_1}(\bar{w_1})
e^{-\bar\phi}\psi^{n_1}(\bar{w_2})>
\cr\times<\lbrack:\psi^5_{t_1}:
(\bar\partial{X^{t_1}}+i(k^{||}\bar\psi^{||})\bar\psi^{t_1})
e^{ik_1^{||}X}+c.c.\rbrack(z_1,\bar{z_1})\cr\times\lbrack
:\psi^5_{t_2}:(\bar\partial{X^{t_2}}+(\bar\psi^{||})\bar\psi^{t_2})
e^{ik_2^{||}X}+c.c.\rbrack(z_2,\bar{z_2})\cr\times
(\partial{X_{m_1}}+i(p_1\psi)\psi_{m_1})e^{ip_1{X}}(w_1,\bar{w_1})
\cr\times
(\partial{X_{m_1}}-i((p_1+k_1^{||}+k_2^{||})\psi)\psi_{m_1})
e^{-i(p_1+k_1^{||}+k_2^{||}){X}}(w_2,\bar{w_2})>}}
where $z_1,z_2$ and $w_2$ are now fixed by the conformal
invariance and will be
 later set to $0,1$ and $\infty$
Evaluation the free theory correlators in (17) gives
\eqn\grav{\eqalign{A(p_1)\cr=
{\lambda^{({+1})}}\lambda^{(-3)}
\int{{d^4{k_1^{||}}}\over{{k_1^{||}}^4}}
\int{{d^4{k_2^{||}}}\over{{k_2^{||}}^4}}
\int{d^2}{w_1}|z_1-w_2|^2|z_2-w_2|^2|z_1-z_2|^2
(\bar{w_1}-\bar{w_2})^{-2}
\cr\times\lbrack\lbrace{{{2{p_1^{t_1}}}{p_2^{t_2}}+\eta^{t_1t_2}
(p_1p_2)^{||} }\over
{(z_1-w_1)^2(z_2-w_2)^2
}}+
{{{2{p_1^{t_2}}}{p_2^{t_1}}+\eta^{t_1t_2}
(p_1^{||}p_2^{||})}\over
{(z_1-w_2)^2(z_2-w_1)^2
}}\cr+{{2p_1^{t_1}p_2^{t_2}+2p_2^{t_1}p_1^{t_2}+\eta^{t_1t_2}}\over
{(z_1-w_1)(z_1-w_2)(z_2-w_2)(z_2-w_1)}}-
{{(8(p_1^{||}k_1^{||})+8(p_1^{||}k_2^{||}))\eta^{t_1t_2}}\over
{{(z_1-z_2)^2}(w_1-w_2)^2}}\cr-{
{\eta^{t_1t_2}(11(p_1^{||}k_1^{||})+11(p_1^{||}k_2^{||})-8(p_1^{||})^{2})}
\over{(z_1-z_2)(w_1-w_2)}}({1\over{(z_1-w_1)(z_2-w_2)}}+
{1\over{(z_1-w_2)(z_2-w_1)}})
\rbrace\cr\times\lbrace
{{{{p_1}_{t_1}}{p_1}_{t_2}}\over{(\bar{z_1}-\bar{w_1})(\bar{z_2}-\bar{w_1})}}
+{{{{p_2}_{t_1}}{p_2}_{t_2}}\over{(\bar{z_1}-\bar{w_2})(\bar{z_2}-\bar{w_2})}}
\cr+
{{{{p_1}_{t_1}}{p_2}_{t_2}}\over{(\bar{z_1}-\bar{w_1})(\bar{z_2}-\bar{w_2})}}
+{{{{p_1}_{t_2}}{p_2}_{t_1}}\over{(\bar{z_1}-\bar{w_2})(\bar{z_2}-\bar{w_1})}}
+{{\eta_{t_1t_2}}\over{(\bar{z_1}-\bar{z_2})^2}}\rbrace+c.c.\rbrack
\cr\times\lbrack
{{{1}\over{|w_1-w_2|^{2(p_1p_2)}|z_1-z_2|^{2(k_1^{||}k_2^{||})}
|z_1-w_1|^{2(k_1^{||}p_1^{||})}}}}
\cr\times
{{1}\over{|z_1-w_2|^{2(k_1^{||}p_2^{||})}
|z_2-w_1|^{2(k_2^{||}p_1^{||})}|z_2-w_2|^{2(k_2^{||}p_2^{||})}}}\rbrack}}
Here $p_{1,2}^{||}$ is the projection of the dilaton momentum on
the four longitudinal directions (parallel to the D3-brane worldvolume).
Setting $z_1\rightarrow{0}$, $z_2\rightarrow{1}$,$w_2\rightarrow\infty$
and evaluating the integral over $w_1$
we find that the $\lambda^2$ contribution to the amplitude
is given by:
\eqn\grav{\eqalign{A(p_1)=
{\lambda^{({+1})}}\lambda^{(-3)}
\int{{d^4{k_1^{||}}}\over{{k_1^{||}}^4}}
\int{{d^4{k_2^{||}}}\over{{k_2^{||}}^4}}\int{d^2}w_1\lbrace\lbrack
({1\over{w_1^2}}+{1\over{\bar{w_1}^2}})(
2p_1^{t_1}p_2^{t_2}+ (p_1^{||}p_2^{||})\eta^{t_1t_2})
\cr+({{1}\over{{1-w_1}^2}}+{{1}\over{{1-\bar{w_1}}^2}})
(2p_1^{t_2}p_2^{t_1}+ (p_1^{||}p_2^{||})\eta^{t_1t_2})
\cr+({{2}\over{w_1(1-w_1)}}+{{2}\over{\bar{w_1}(1-\bar{w_1})}})
(p_1^{t_1}p_2^{t_2}+p_1^{t_2}p_2^{t_1}+ (p_1^{||}p_2^{||})\eta^{t_1t_2})
\cr\times\rbrack
p_1^{t_1}p_1^{t_2}|w_1|^{2(k_1^{||}p_1^{||})}|1-w_1|^{2(k_2^{||}p_1^{||})}
\rbrace
\cr=2\int{{d^4{k_1^{||}}}\over{{k_1^{||}}^4}}
\int{{d^4{k_2^{||}}}\over{{k_2^{||}}^4}}\lbrace
\lbrack2(p_1^{||})^4+(p_1^{||})^2(p_1^{||}k_1^{||})
+(p_1^{||})^2(p_1^{||}k_2^{||})\rbrack\cr\times
{{\Gamma(1+(k_1^{||}p_1^{||}))\Gamma((p_1^{||}k_2^{||})-1)
\Gamma(1-(p_1^{||}(k_1^{||}+k_2^{||})))}\over
{\Gamma((p_1^{||}(k_1^{||}+k_2^{||})))\Gamma((-k_1^{||}p_1^{||}))
\Gamma(2-(k_2^{||}p_1^{||}))}}\rbrace}}
where $\Gamma$ is Euler gamma-function.
It is remarkable that this amplitude depends exclusively
on the longitudinal projection of the  dilaton momentum
$p_1$,i.e. the four-dimensional vector $p_1^{||}$.
That means that , upon the Fourrier transform
$A(p_1)$ will be the function of four space-time coordinates,
corresponding to the $AdS_5$ boundary (or the polarization
of the 5-form vertex).We observe
 that the structure of the two-dilaton amplitude
involves the fourth power of the four-dimensional momentum
$p_1^{||}$ (appearing as as the kinematic factor
in the NSR superstring four-point function) 
multiplied by the factor that will arise as a result
of the  integration of the product of gamma-functions
in the Veneziano amplitude over the momenta. We will
show that this factor is proportional to the $\sim{ln({p_1^{||}}^2)}$,
i.e. $A(p_1)\sim{{{p_1^{||}}^{4}}}ln({p_1^{||}}^2)$
  indeed reproduces
the two-point correlation function $<F^2(p)F^2(-p)>$ in the 
four-dimensional $N=4$ super Yang-Mills.
Our aim now is to perform the integration over
$k_1^{||}$ and $k_2^{||}$ in the amplitude (19).
Let us cast the expression (19) for the dilaton 
amplitude into the following form:
\eqn\grav{\eqalign{
A(p_1)=
2{\lambda^{({+1})}}\lambda^{(-3)}(p_1^{||})^2
\int{{{d^2}{w_1}}\over{|1-w|^4}}
\int{{d^4{k_1^{||}}}\over{{k_1^{||}}^4}}
\int{{d^4{k_2^{||}}}\over{{k_2^{||}}^4}}
\lbrack{2}(p_1^{||})^2+(p_1^{||}(k_1^{||}+k_2^{||}))\rbrack\cr\times
e^{2i(k_1^{||}p_1^{||})ln|w_1|+(k_2^{||}p_1^{||})ln|1-w_1|}
}}
We start with the term proportional to ${p_1^{||}}^4$ in (20).
We need to evaluate the integral over $k_1^{||}$
(the integration over $k_2^{||}$ is totally similar).
To make our notations more convenient, we denote
$k_1^{||}\equiv{k}$ and $p_1^{||}\equiv{p}$ and write:
\eqn\grav{\eqalign{I_1(p_1,w_1)=\int{{d^4{k}}\over{{k}^4}}
e^{2i(kp)ln|w|}
=\int{d^3}{{\vec{k}}}{\int_{-\infty}^{\infty}}d{k_0}
{{e^{2i({k_0p_0}-({\vec{k}}{\vec{p}}))ln|w|}}\over
{(k_0-|{\vec{k}}|+i\epsilon)^2(k_0+|{\vec{k}}|-i\epsilon)^2}}
}}
Integrating over $k_0$ first by evaluating the residues we 
get the spatial integral over ${\vec{k}}$ which is not difficult
to compute:

\eqn\grav{\eqalign{I_1(p_1,w_1)=-\int{d^3}{\vec{k}}
\partial_{|{{k}}|}{{e^{2i|{{k}}|(p_0-|{{p}}|cos{\theta})ln|w|}}
\over{4|{{k}}^2|}}=
{\int_0^\infty}d|k||k|^2\int_0^{\pi}d\theta{sin}\theta\cr\times
\lbrack{{e^{2i|k|(p_0-p{cos}\theta)ln|w|}}\over{2|k|^3}}-
{{2i(p_0-p{cos}\theta)ln|w|e^{2i|k|(p_0-|p|cos\theta)ln|w|}}\over{4|k|^2}}
\rbrack\cr={\int_0^{\infty}}d|k|
(e^{2i(p_0+|p|)|k|ln|w|}-e^{2i(p_0-|p|)|k|ln|w|})\cr\times
({1\over{4|k|}}{1\over{8i|p||k|^2ln|w|}}(1-{p_0\over{|p|}})-
{1\over{8ip|k|^2ln|w|}})\cr=
-{1\over4|p|}(p_0+|p|)(ln\lbrack|2ln|w|(p_0+|p|)|\rbrack-1)
\cr+{1\over4|p|}(p_0-|p|)(ln\lbrack|2ln|w|(p_0-|p|)|\rbrack-1)
\cr-{1\over4}
(1-{{p_0}\over{|p|}})ln(p_0+|p|)-{1\over4}
(1+{{p_0}\over{|p|}})ln(p_0-|p|)\cr=
{1\over2}(1+ln|2ln|w||-ln(p_0-|p|)-ln(p_0+|p|))=
{1\over2}(1-ln|2ln|w||-{1\over2}ln(p^2))}}

where we  introduced $|p|\equiv|{\vec{p}}|$ and used that
the regularized  integrals are given by:
$\int_0^\infty{{e^{i\alpha{x}}}\over{x}}dx=-ln{\alpha}$, 
$\int_0^\infty{{e^{i\alpha{x}}}}dx=-{1\over{i\alpha}}$ etc.
The similar computation shows that
\eqn\grav{\eqalign{\int{{d^4{k_1^{||}}}\over{{k_1^{||}}^4}}
\int{{d^4{k_2^{||}}}\over{{k_2^{||}}^4}}
(p_1^{||}(k_1^{||}+k_2^{||}))
e^{2i(k_1^{||}p_1^{||})ln|w_1|+(k_2^{||}p_1^{||})ln|1-w|}\cr
={1\over{8ln|w|}}(-{1\over2}ln({p_1^{||}}^2)-ln(|2ln|w||)+3)
(1-ln(|2ln|1-w||)-{1\over2}ln({p_1^{||}}^2))\cr+
{1\over{8ln|1-w|}}(-{1\over2}ln({p_1^{||}}^2)-ln(|2ln|1-w||)+3)
(1-ln(|2ln|w||)-{1\over2}ln({p_1^{||}}^2))}}
Using (20), (22), (23), the expression for the two-point
dilaton amplitude is given by:
\eqn\grav{\eqalign{A(p_1)=
 2{\lambda^{({+1})}}\lambda^{(-3)}
\int{{{d^2}{w_1}}\over{|1-w|^4}}\lbrace
(p_1^{||})^4(1-ln(|2ln|w||)-{1\over2}ln((p_1^{||})^2))
\cr{\times}(1-ln(|2ln|1-w||)-{1\over2}ln((p_1^{||})^2))\cr+
{{({p_1^{||}})^2}\over{8ln|1-w|}}(-{1\over2}ln({p_1^{||}}^2)-ln(|2ln|1-w||)+3)
(1-ln(|2ln|w||)-{1\over2}ln({p_1^{||}}^2))\rbrace}}
For a time being, we shall concentrate on the part proportional to
$\sim({p_1^{||}})^4$ in the amplitude $A(p_1)$
 as it clearly contributes to the most singular
part in the correlator $<F^2(x)F^2(y)>$ upon the Fourier transform.
Since 
\eqn\lowen{\int{d^2}w{{1}\over{|w|^\alpha|1-w|^\beta}}=
{{{\Gamma(1-{\alpha\over2})}{\Gamma(1-{\beta\over2})}
{\Gamma({{\alpha+\beta}\over2}-1)}}\over{\Gamma(2-{{\alpha+\beta}\over2})
\Gamma({\alpha\over2})\Gamma({\beta\over2})}}}
one has 
\eqn\lowen{\int{{{d^2}{w_1}}\over{|1-w|^4}}=
{{\Gamma(-1)\Gamma(1)\Gamma(1)}\over{\Gamma(2)\Gamma(0)\Gamma(0)}}=
{1\over{\Gamma(0)}}=0}
therefore the term proportional to the square of the logarithm
$\sim({ln}{(p_1^{||})^2})^2$ vanishes.
Then the leading non-analytic term in the amplitude is given by:
\eqn\grav{\eqalign{A(p_1){\sim}2{\lambda^{({+1})}}\lambda^{(-3)}
{(p_1^{||})^4}ln{(p_1^{||})^2}\int{d^2}{w}
{{ln(|ln|w||)-ln(|ln|1-w||)}\over{|1-w|^4}}}}
Fourier transforming (27) to the four-dimensional position space we get
\eqn\grav{\eqalign{<F^2(x)F^2(y)>\equiv<V_\varphi{V_\varphi}>\sim
{{{}{\lambda^{({+1})}}\lambda^{(-3)}}\over{|x-y|^8}}+ ...}}
where we have skipped less singular terms, possibly arising
from the $(p_1^{||})^2$-contribution.
 As we mentioned before, the dependence on
N is actually hidden in  the factor of $\lambda$.
Since the two-point correlator is proportional to $N^2$ 
one has to impose the condition ${\lambda^{({+1})}}\lambda^{(-3)}\sim{N^2}$.
We will choose

\eqn\grav{\eqalign
{\lambda^{(+1)}\sim{{\rho}\over{{N}}}\cr
\lambda^{(-3)}\sim{{\rho}N^3}}}

where $\rho$ is N-independent constant chosen in such a way that
the product of $\rho^2$ with the regularized integral
over $w$ in (27) gives the correct normalization of the two-point
function in the four-dimensional super Yang-Mills theory.
Then we have
\eqn\lowen{<F^2(x)F^2(y)>\sim{{N^2}\over{|x-y|^8}}}
At the first glance
the choice (29) of N-dependence of ${\lambda^{(+1)}}$ and
${\lambda^{(-3)}}$ in (29) is  not unique
(since it is only important that their product is proportional 
to $N^2$). However we shall see that due to the ghost number
conservation
 the $N$-dependence choice
(29) is actually uniquely fixed by 
higher order corrections in $\lambda^2$ which we  will discuss
in the next section.

\centerline {\bf Corrections in $\lambda$ as partial waves.}
In this part we will make few comments on the structure 
of amplitudes arising as a result of further expansion in $\lambda^2$
the contributions of order $\lambda^{2n}$ correspond to
$2n+2$-point correlation functions integrated over $2n$
internal momenta of the five-form vertices.
The kinematic factor correlation functions  is some polynomial
of degree $2n+2$ while the integration over the internal momenta
should hopefully produce the logarithm 
similarly to the four-point case so the
most singular part of the $\lambda^{2k}$ contribution
to the dilaton-dilaton amplitude in the model (13) is proportional
to $\sim{k^{2n+2}}ln{k^2}$ In the AdS picture 
such a momentum dependence corresponds to the
contribution of the dilaton partial wave with the angular momentum
$l=2n-2$. Therefore  the expansion in $\lambda^2$ 
in (13) seems to correspond
to the expansion in  partial waves of the dilaton.
Also the comment should be made about the N-dependence of the 
dilaton-dilaton amplitude.
The choice (29) for the N-dependence insures the proper 
N-dependence for the 4n-point amplitudes (proportional to $\lambda^{4n-2}$),
i.e. the partial waves with even angular momenta.
As for the waves with odd angular momenta (corresponding to
$4n+2$-point amplitudes) in the sigma-model, in order to take them into
account with the proper weight one has to insert the picture-changing factor 
in the measure of functional integration so 
as to insure  the odd momentum
contributions are correctly weighted. Namely, one has to choose 
the picture-changing function $f(\Gamma)$ in the
  generating functional (13) as
\eqn\lowen{ f(\Gamma)=1+{\Gamma^2\over{N^2}}}
 provided  that the parameters $\lambda^{(-3)}$ and $\lambda^{(+1)}$
 are defined 
as in (29). As we already mentioned before,
there exists also an equivalent way of writing the generating functional
(13) with only the picture $+1$
version of the five-form present in the action:
\eqn\grav{\eqalign{Z(\varphi,\lambda)=
\int{D}\lbrack{X}\rbrack{D}\lbrack\Psi\rbrack
\lbrack{ghosts}\rbrack {{\Gamma^2}\over{\Gamma^2-N^2}}
{exp}{\lbrace}\int{d^2{z}}
{1\over2}(\partial{X_m}\bar\partial{X^m}+\psi_m\bar\partial\psi^m)
\cr+\int{{{d^4}k^{||}}\over{{k^{||}}^4}}\epsilon^{p_1...p_4}
{\rho\over{N}}e^{\phi}
\psi_{p_1}...\psi_{p_4}\psi_t(\bar\partial{X^t}+
i(k^{||}\psi)\psi^t)e^{ik^{||}x^{}}+cc
\cr+\int{d^{10}k}V_\varphi(k)\varphi(k)\rbrace}}
It is easy to show that the functional (32) and the functional
(13) with the measure function (31) and with  
$\lambda^{(-3)}$ and $\lambda^{(+1)}$ defined in (29)
 produce equivalent
series  in $\rho$.

\centerline{\bf Conclusion}
We have shown that the dilaton-dilaton amplitude 
in the ten-dimensional branelike sigma-model (13), (32)
reproduces the two-point correlation function $<F^2F^2>$
of the $N=4$ $D=4$ super Yang-Mills theory.
The natural project for the future is to compute 
three  and four-point correlators in
 the branelike sigma-model model (13), (32) in order
to verify their agreement with  the corresponding
computations in the AdS supergravity (see, for instance,
~\refs{\hokerf, \hokers, \hokert, \seiberg, \bianchi})
The questions of
particular interest would be reproducing the logarithmic
singularities in  four-point correlators in the AdS supergravity
 and the relation between logarithms in AdS correlators
and anomalous dimensions of Yang-Mills operators.
Another important object of interest is AdS S-matrix 
~\refs{\pol, \pols}
which can be understood in terms of structure constants
(or three-point correlation functions) of the conformal field theory
on the worldsheet of superstring on $AdS_5\times{S^5}$.
In accordance with our model these functions are to be computed in 
the free theory with five-form insertions.
The final observation to be made is that the fiveform 
state in the spectrum of NSR superstring in $D=10$  seems to play 
the crucial role in building the space-time geometry.
Namely, the five-form vertex $V_5$
(which defines a BRST-invariant massless state in superstring theory in
$flat$ ten-dimensional space-time)
 integrated over its four-dimensional
momentum , transforms the flat maximally supersymmetric
space-time vacuum into the one of $AdS_5\times{S^5}$.
Thus the Anti-de-Sitter structure of our space-time appears to have 
 dynamical origin: it is the consequence of the presence
of exotic brane-like states in the spectrum of a superstring.
We hope that the computations outlined above
will strengthen the ground for this hypothesis.
\centerline{\bf Acknowledgements}
The author gratefully acknowledges the hospitality and support of the
Abdus Salam International Centre for Theoretical Physics (ICTP)
in Trieste and useful discussions with M.Bianchi, K.S.Narain
and  A.M.Polyakov.
\listrefs
\end